\documentclass[runningheads]{llncs}

\usepackage[T1]{fontenc}
\usepackage{graphicx}
\usepackage{url}
\usepackage{hyperref}
\usepackage{textcomp}
\usepackage{float}

\usepackage{amssymb} 
\usepackage{xcolor}  

\newcommand{\cmark}{\textcolor[RGB]{0,128,0}{\checkmark}} 
\newcommand{\xmark}{\textcolor[RGB]{128,0,0}{\(\times\)}} 

\long\def\comment#1{}

\hypersetup{
    colorlinks=true,
    linkcolor=blue,
    filecolor=magenta,      
    urlcolor=cyan,
    citecolor=blue
}

\begin{document}

\title{AmpliconHunter: A Scalable Tool for PCR Amplicon Prediction from Microbiome Samples}
\titlerunning{AmpliconHunter}
%

\author{Rye Howard-Stone\inst{1} \and Ion I. M\u{a}ndoiu\inst{1}\orcidID{0000-0002-4818-0237}}

\authorrunning{R. Howard-Stone and I.I. M\u{a}ndoiu}

\institute{School of Computing, University of Connecticut, Storrs CT 06269, USA\\
\email{\{rye.howard-stone,ion.mandoiu\}@uconn.edu}}

\maketitle

\begin{abstract}
Sequencing of PCR amplicons generated using degenerate primers (typically targeting a region of the 16S ribosomal gene) is widely used in metagenomics to profile the taxonomic composition of complex microbial samples.  To reduce taxonomic biases in primer selection it is important to conduct \emph{in silico} PCR analyses of the primers against large collections of up to millions of bacterial genomes. However, existing \emph{in silico} PCR tools have impractical  running time for analyses of this scale.  In this paper we introduce AmpliconHunter, a highly scalable \emph{in silico} PCR package distributed as open-source command-line tool and publicly available through a user-friendly web interface at \href{https://ah1.engr.uconn.edu/}{https://ah1.engr.uconn.edu/}.  AmpliconHunter implements an accurate nearest-neighbor model for melting temperature calculations, allowing for primer-template hybridization with mismatches, along with three complementary methods for estimating off-target amplification.  By taking advantage of multi-core parallelism and SIMD operations available on modern CPUs, the AmpliconHunter web server can complete \emph{in silico} PCR analyses of commonly used degenerate primer pairs against the 2.4M genomes in the latest AllTheBacteria collection in as few as 6-7 hours. 
\end{abstract}

\section{Introduction}

DNA amplification by Polymerase Chain Reaction (PCR) is a fundamental technique in biomedical research. Commonly, a genomic region of interest is targeted for amplification by identifying a pair of primer sequences that flank the region.  Sequencing PCR amplicons (typically variable regions of the 16S ribosomal gene) remains a very popular approach in metagenomic studies, where amplicon sequencing has been shown to enable identification of more taxa at a lower cost compared to shotgun sequencing \cite{Gehrig2022}. To ensure sensitive amplification of the target region even with high-levels of genomic variation in bacterial species, it is common to use degenerate primers and permissive PCR conditions that allow primers to anneal with mismatches \cite{AbellanSchneyder2021}. Degenerate primers are represented as sequences over the IUPAC alphabet, in which each letter indicates a subset of one or more nucleotides (e.g., `N' represents any DNA base, `R' represents `A or G', etc.). Experimentally, a degenerate primer is a mixture consisting of all oligonucleotides that can be formed by selecting compatible DNA bases at each degenerate position of the primer sequence. The degree of degeneracy for commonly used primers varies, with some degenerate primers consisting of hundreds of distinct oligonucleotides.
For example, the two degenerate primers targeting the  Titan\texttrademark \space region of the ribosomal operon \space (previously known as StrainID\texttrademark \space \cite{Graf2021}) together consist of a total of 960 oligonucleotides.

Primer selection for amplicon metagenomic sequencing involves complex tradeoffs between sensitivity (which increases with primer degeneracy), specificity (which decreases with degeneracy), taxonomic discriminative power (which depends on the length and diversity of the amplified region), and cost (long amplicons require the use of more expensive sequencing technologies), among others. Most existing primer sequences are designed based on highly curated microbial genome sequences available in databases such as the reference genomes in RefSeq \cite{Ye2012,OLeary2016RefSeq}. This may bias resulting primers towards culturable microbes \cite{Hug2018}. The bias could be reduced by designing and validating the primers \emph{in silico} against larger genome collections that are now becoming available, such as the Genome Taxonomy Database (GTDB, currently containing over 580K genomes \cite{Parks2021GTDB}), the Pathosystems Resource Integration Center database (PATRIC, with nearly 1M genomes) \cite{Gillespie2011PATRIC}, or the AllTheBacteria project (with over 2.4M genomes \cite{Hunt2024AllTheBacteria}). Since analysis of genomic collections of this size is impractical with existing tools there is a need for more scalable and accurate \emph{in-silico} PCR software packages. 

In this paper, we present AmpliconHunter, a highly scalable open-source \emph{in silico} PCR package available at \href{https://github.com/rhowardstone/AmpliconHunter}{github.com/rhowardstone/AmpliconHunter}.  
The AmpliconHunter program accepts as input genome sequences in FASTA format, degenerate primer sequences, and PCR parameters such as annealing temperature and allowed number of mismatches and reports detailed information on the predicted PCR amplicons, including their sequences, genomic coordinates, and predicted melting temperatures for primer binding sites.  To facilitate the use of AmpliconHunter by users without command-line expertise or access to high-performance computing infrastructure, the tool is also publicly available as a web server at \href{https://ah1.engr.uconn.edu/}{https://ah1.engr.uconn.edu/}. The web server can be used to analyze any user specified primer pairs against several pre-loaded bacterial collections.

AmpliconHunter is optimized for the analysis of large genome collections, with the webserver predicting amplicons for the 2.4M AllTheBacteria genomes in 6.5-7 hours, depending on primer degeneracy.  Amplification predictions are based on accurate primer-target melting temperature estimates using a nearest-neighbor model that allows mismatches \cite{Cock2009}.  A distinguishing feature of AmpliconHunter is that it includes multiple methods for estimating the amount of off-target amplification using the primer hybridization annotations, scoring for homology using a profile HMM trained on-the-fly using high-confidence target amplicons, as well as analyses based on decoys generated by reversing the genome sequences.
Such off-target amplification becomes an increasingly important concern when using highly degenerate primer pairs and permissive PCR annealing conditions, particularly for highly complex microbial samples. As shown in Appendix \ref{appendix:offtarget}, off-target amplification predicted by AmpliconHunter is detectable in PacBio HiFi reads generated by sequencing Titan amplicons from a wild mouse gut microbiome sample. 

In the rest of the paper we describe the methods used by AmpliconHunter in Section \ref{sec.methods}, present empirical evaluation results in Section \ref{sec.results}, and conclude with directions for future work in Section \ref{sec.conclusions}.

\section{Methods}
\label{sec.methods}

\subsection{AmpliconHunter Workflow}

\begin{figure}[t]
    \centering
    \includegraphics[width=0.4\linewidth]{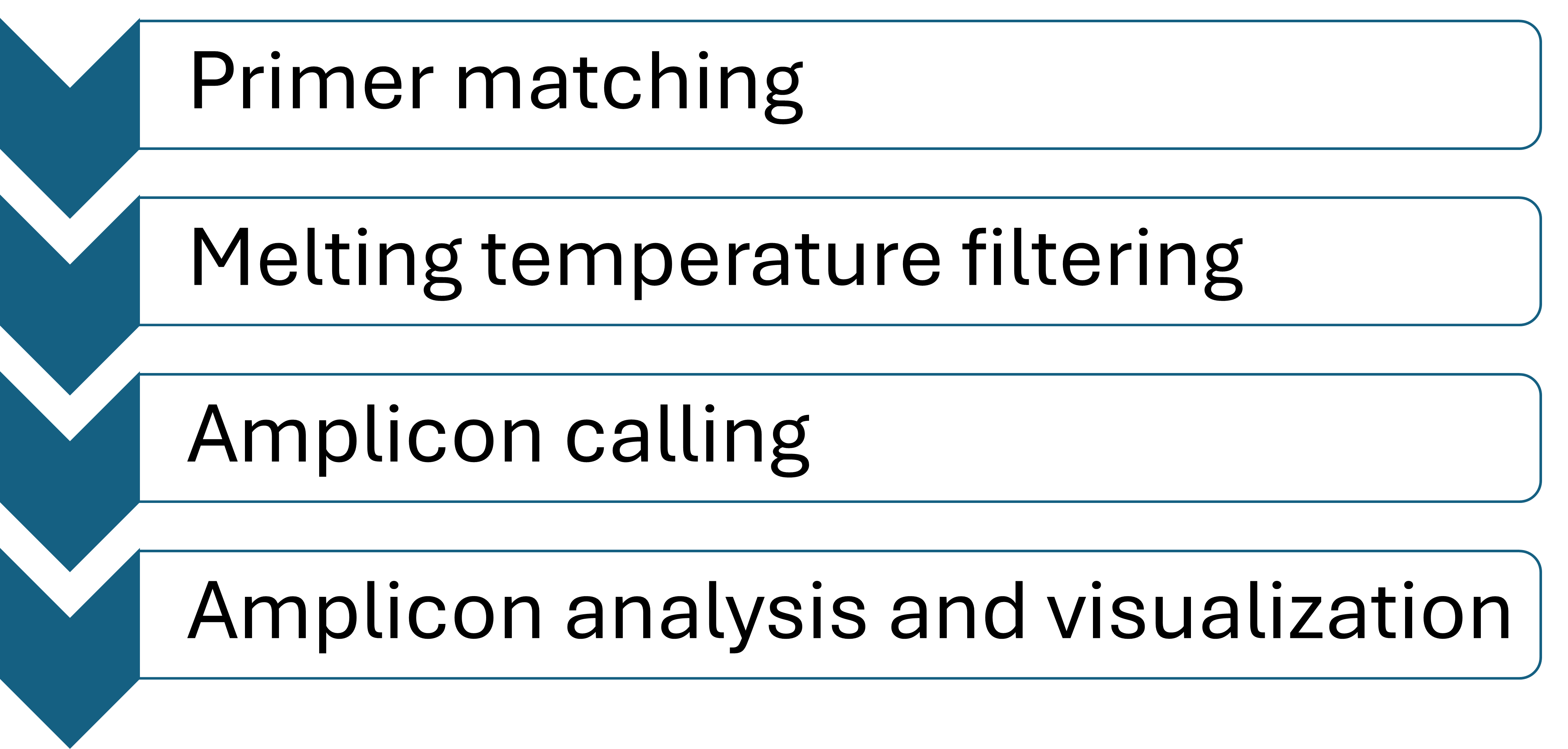}
    \caption{Main steps of the AmpliconHunter workflow.}
    \label{fig:enter-label}
\end{figure}

Figure \ref{fig:enter-label} gives a high-level overview of AmpliconHunter's workflow. The main steps are individually discussed below.

\subsubsection{Primer matching.}
The core pattern matching step in AmpliconHunter generates a list of candidate primer hybridization sites using Intel’s Hyperscan high-performance regular expression matching library, which takes advantage of SIMD capabilities of modern CPUs \cite{Wang2019}. In this step, degenerate primers are converted to regular expressions that account for all possible oligonucleotide variants according to the IUPAC nucleotide codes and the maximum allowed number of mismatches. An optional, variable length clamp region may be specified by the user, such that no mismatches are allowed in the 3’-most bases of each primer match. Clamp filtering is conducted as a post-processing step after regular expression matching and prior to melting temperature calculations.

\subsubsection{Melting temperature filtering.}
AmpliconHunter uses BioPython’s nearest-neighbor model for DNA duplex stability to calculate the melting temperature of each candidate binding site identified in the previous step. Advanced users may specify detailed PCR parameters, including the concentrations of the primer (dnac1), template (dnac2), along with the concentrations of various ions (Na, K, Tris, Mg, and dNTPs), and salt correction method \cite{Cock2009}.
Users can also specify a minimum melting temperature threshold. This threshold is applied to the \emph{maximum} melting temperature over all oligonucleotides represented by the primer, i.e., a candidate binding site is retained if at least one of the oligonucleotides represented by the primer has a melting temperature above the threshold. 
Unlike other libraries such as Python's Regex library, Hyperscan does not return the mismatch positions for fuzzy matching, thus all compatible oligonucleotides must be checked individually during melting temperature filtering. 
The maximum melting temperatures for retained binding sites are saved and included in amplicon header annotations for the sites predicted to yield exponential amplification by the amplicon calling step.

\subsubsection{Amplicon calling.}
After primer annealing sites have been filtered for 3' clamp and melting temperature considerations, candidate amplicons are generated by pairing primer hybridization sites found on opposite strands and within the user-specified range of amplicon lengths.  While shorter amplicons are likely to be amplified with greater efficiency compared to longer amplicons, AmpliconHunter does not seek to predict amplicon yield and instead reports all genome regions that may be amplified and detected after amplicon sequencing.

\subsubsection{Amplicon analysis.}
AmpliconHunter implements three complementary methods for assessing potential off-target amplification, a feature often neglected by other \emph{in silico} PCR programs:
\begin{enumerate}
    \item Amplicons are annotated according to their primer pairing - those formed from two of the same type of primer are found and annotated as either ‘FF’ or ‘RR’ orientation (versus forward strand amplicons ‘FR’, or reverse ‘RF’). Please see Appendix \ref{appendix:offtarget} for more information.
    \item A profile HMM is trained on-the-fly with high-confidence amplicons (exact primer matches only) obtained  with the same primers on the RefSeq-complete dataset, then used to score all predicted amplicons called in the previous step.
    \item An empirical estimate of the off-target amplification rate is computed by calling amplicons from decoys generated by reversing the given genome sequences (this is an optional feature on the AmpliconHunter web server since performing decoy analysis doubles the required time). 
\end{enumerate}

\subsubsection{Visualization.}
The main AmpliconHunter output is a fasta file with predicted amplicon sequences, their genomic coordinates, primer orientations, and maximum annealing temperatures.
Additionally, the program generates visualizations of length and orientation statistics. When taxonomy information is provided the program also generates heatmaps of amplicon similarity across species within each genus.
The web interface further extends these capabilities with interactive visualizations and the ability to compare results across different parameter settings. All visualizations are available for download.

\subsection{Performance Optimization}
The tool utilizes Python’s multiprocessing library to parallelize searches across available CPU cores. The web server implements a caching system for both profile HMM models and amplicon calling results, using content-addressable storage based on input parameter hashes. By caching past runs the web server response time becomes negligible when the same primer pair has been analyzed before with the same search parameters.

To guard against combinatorial explosion of runtime and output size, the server imposes mild limitations on user-specified parameters. No primer may have degeneracy of more than 10,000, the maximum number of mismatches must be at most 10, and each primer must be at least 5bp in length. If the users wish to conduct \emph{in silico} PCR analysis of custom collections of genomes or use parameters outside these ranges, they must download and run the command-line version of AmpliconHunter locally.

\subsection{Benchmarking Setup}
Five databases \cite{OLeary2016RefSeq,Parks2021GTDB,Gillespie2011PATRIC,Hunt2024AllTheBacteria} ranging in size from 5.7K to 2.4M genomes are available through the web interface, each with taxonomy to the species level for many entries (see \autoref{tab:database_summary}). These options allow the users to select an appropriate database based on their research needs, balancing between higher quality assemblies (e.g., RefSeq-Complete) and broader taxonomic coverage (e.g., AllTheBacteria).

\begin{table}[tbp]
\centering
\caption{Summary of databases available on the AmpliconHunter web server.}
\begin{tabular}{|l|c|c|p{6cm}|}
\hline
\textbf{Database} & \textbf{Genomes} & \textbf{Citation} & \textbf{Description} \\
\hline
RefSeq-Complete & 5,730 & \cite{OLeary2016RefSeq} & A curated collection of complete reference genomes for bacteria \\
\hline
RefSeq-All & 20,578 & \cite{OLeary2016RefSeq} & All genomes in RefSeq-Complete, as well as the incomplete bacterial reference genomes \\
\hline
GTDB & 581,552 & \cite{Parks2021GTDB} & The Genome Taxonomy Database provides a standardized microbial taxonomy based on genome phylogeny. \\
\hline
PATRIC & 989,844 & \cite{Gillespie2011PATRIC} & The Pathosystems Resource Integration Center, a comprehensive genomic database for bacteria. \\
\hline
AllTheBacteria & 2,440,377 & \cite{Hunt2024AllTheBacteria} & A very large dataset of uniformly assembled bacterial genomes \\
\hline
\end{tabular}
\label{tab:database_summary}
\end{table}

The RefSeq-complete database was obtained from NCBI using the Datasets command-line tool (version 17.1.0) with the following command:

\begin{verbatim}
datasets download genome taxon 2 \
  --reference \
  --assembly-level complete \
  --include genome \
  --filename bacteria-complete-refseq.zip
\end{verbatim}

The -\--reference flag was included to restrict results to a small number of high-quality, curated reference genomes from the RefSeq database. Without this flag, the dataset would include all complete genomes, a much larger set ({\raise.17ex\hbox{$\scriptstyle\sim$}}114,000 genomes at the time of writing). The resulting RefSeq-complete database used in this study contained 5,730 genomes.

We conducted benchmarking experiments with four primer pairs targeting different regions of the ribosomal operon:

\begin{itemize}
    \item V3V4 primers (forward: CCTACGGGNGGCNGCAG, reverse: GACTACNNGGGTATCTAATCC), targeting the V3-V4 subregion of the 16S rRNA gene, with an expected amplicon length of 450bp \cite{Walker2015Jun},
    \item V1V9 primers (forward: AGRGTTYGATYMTGGCTCAG, reverse: RGYTACCTTGTTACGACTT), targeting the 16S rRNA gene with expected amplicon length of 1600bp \cite{Mason2009Feb},
    \item Titan primers (forward: AGRRTTYGATYHTDGYTYAG, reverse: YCNTTCCYTYDYRGTACT), targeting the 16S-23S rRNA region with expected amplicon length of 2455bp \cite{Gehrig2022}, and
    \item mecA primers (forward: GTAGAAATGACTGAACGTCCGATAA, reverse: CCAATTCCACATTGTTTCGGTCTAA) targeting the mecA antibiotic resistance gene with expected amplicon length of 310bp \cite{Hosseini2016,Gorecki2019}
\end{itemize}

We assessed AmpliconHunter for both accuracy and computational performance. Tests were conducted on a Ubuntu 22.04 KVM virtual machine configured with 190 virtual cores and 960GB of RAM running on a Dell PowerEdge R7525 server with two AMD EPYC 7552 48-Core Processors and 2TB total RAM.  We compared the Hyperscan version of AmpliconHunter (AHv1) to two recently published \emph{in silico} PCR tools, RibDif2 \cite{Murphy2023} and PrimerEvalPy \cite{VazquezGonzalez2024}, as well as a preliminary prototype implementation of AmpliconHunter using Python's Regex library, referred to as AHv0. 
\comment{Other implementations exist, often optimized as a graphical interface hosted on a web server rather than a command-line program \cite{Kent2002}.} A feature comparison for the compared tools is given in \autoref{tab:feature_comparison}.

\begin{table}[tbp]
\centering
\caption{Feature comparison of compared in-silico PCR tools.}
\begin{tabular}{|l|c|c|c|}
\hline
\textbf{Feature} & \textbf{RibDif2} & \textbf{PrimerEvalPy} & \textbf{AmpliconHunter} \\
\hline
Degenerate primers & \cmark & \cmark & \cmark \\
\hline
Melting temperature & \xmark & \xmark & \cmark \\
\hline
Mismatch parameter & \xmark & \cmark & \cmark \\
\hline
Web server available & \cmark & \xmark & \cmark \\
\hline
Command line interface & \cmark & \cmark & \cmark \\
\hline
Parallel processing & \cmark & \xmark & \cmark \\
\hline
Off-target amplification & \cmark & \xmark & \cmark \\
\hline
Programming language & Python/Perl & Python & Python \\
\hline
\end{tabular}
\label{tab:feature_comparison}
\end{table}

Unless otherwise specified, all methods were run with a time limit of one hour. Accuracy tests were performed using a random subset of 20\% of the genomes in the RefSeq-Complete database and timing experiments were performed using random subsets of the AllTheBacteria database.

\section{Results}
\label{sec.results}

\subsection{Accuracy Comparison}
Our primary reference dataset consists of all complete bacterial genomes from RefSeq, with corresponding taxonomy and rRNA gene annotations. In order to assess accuracy of all methods, we limited input to a random subset consisting of 20\% of these genomes.  For all compared methods, predicted amplicons were labeled as true positive amplicons if the primer binding sites (amplicon ends) fall within the expected gene annotation and maintain the correct strand orientation.

\begin{figure}
    \centering
    \includegraphics[width=1\linewidth]{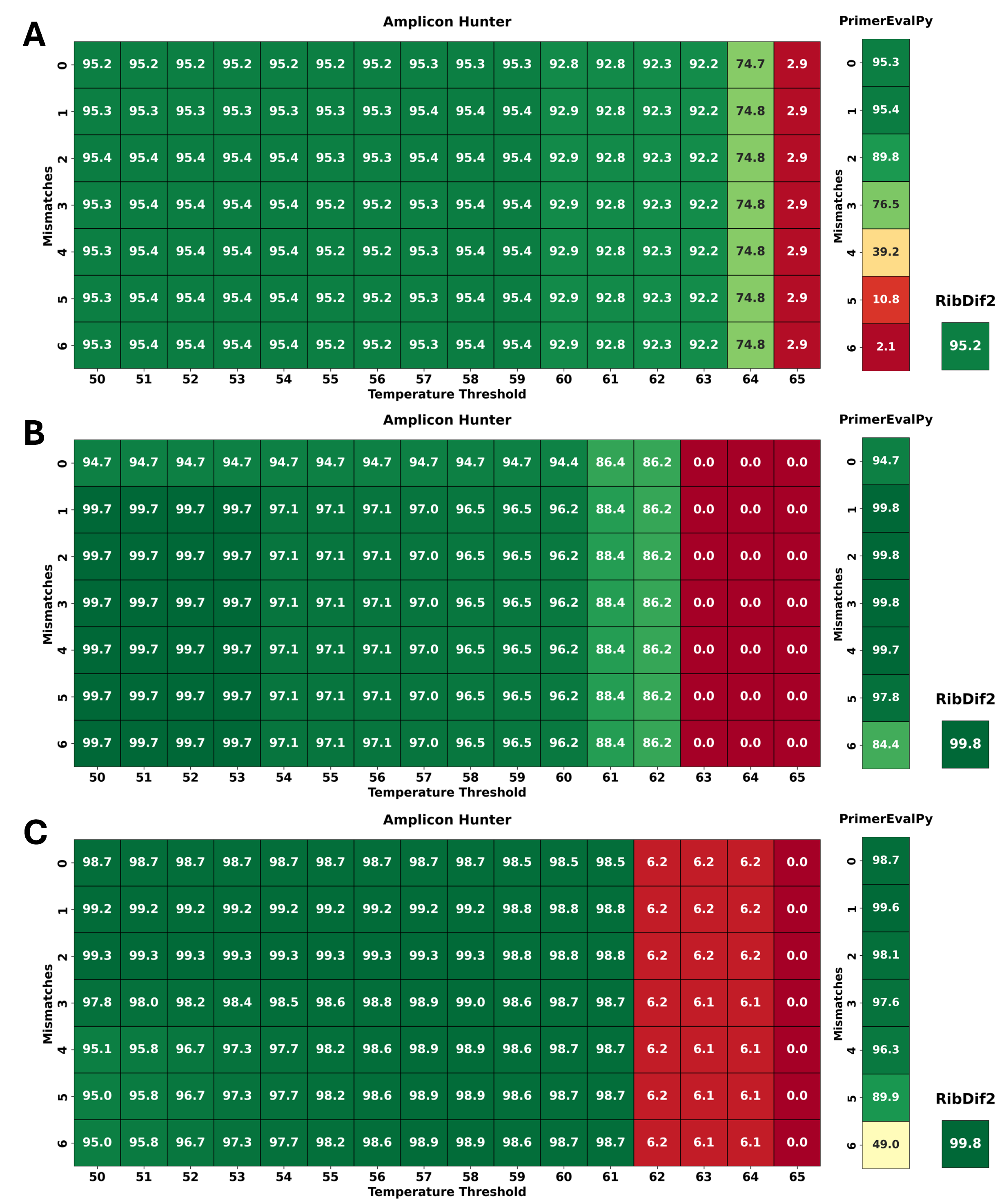}
    \caption{F1 scores for all methods at various parameter settings on a subset of the RefSeq-Complete database, using three primers pairs A) Titan, B) V1-V9, and C) V3V4 primers.}
    \label{fig:accuracy}
\end{figure}

Figure \ref{fig:accuracy} gives the F1 scores (defined as the harmonic mean of precision and recall) obtained by running the compared tools using the Titan, V1V9, and V3V4 primer pairs. Whenever possible, the tools were run with varying minimum melting temperatures and maximum number of mismatches (PrimerEvalPy and RibDif2 do not model melting temperatures, and RibDif2 uses a hard-coded threshold of up to 1 mismatch and 1 indel).  All three tools achieve comparable F1 maximum scores for all three primer pairs. AmpliconHunter's accuracy is less sensitive to varying the maximum numbers of mismatches than PrimerEvalPy due to its additional filtering based on melting temperatures. This difference is most pronounced when using highly degenerate primers such as the Titan pair.

\subsection{Runtime Comparison}
We conducted three scaling tests to evaluate computational efficiency. All tests used the V1V9 primer pair with default parameters (2 mismatches allowed, 50°C minimum annealing temperature) unless otherwise specified. A test run was terminated if it did not complete with one hour.  In all comparisons, 
AmpliconHunter dramatically outperforms all other methods, including the preliminary prototype implementation using Python's Regex library.

\subsubsection{Input size scaling.}
\begin{figure}[tb]
    \centering
    \includegraphics[width=0.5\linewidth]{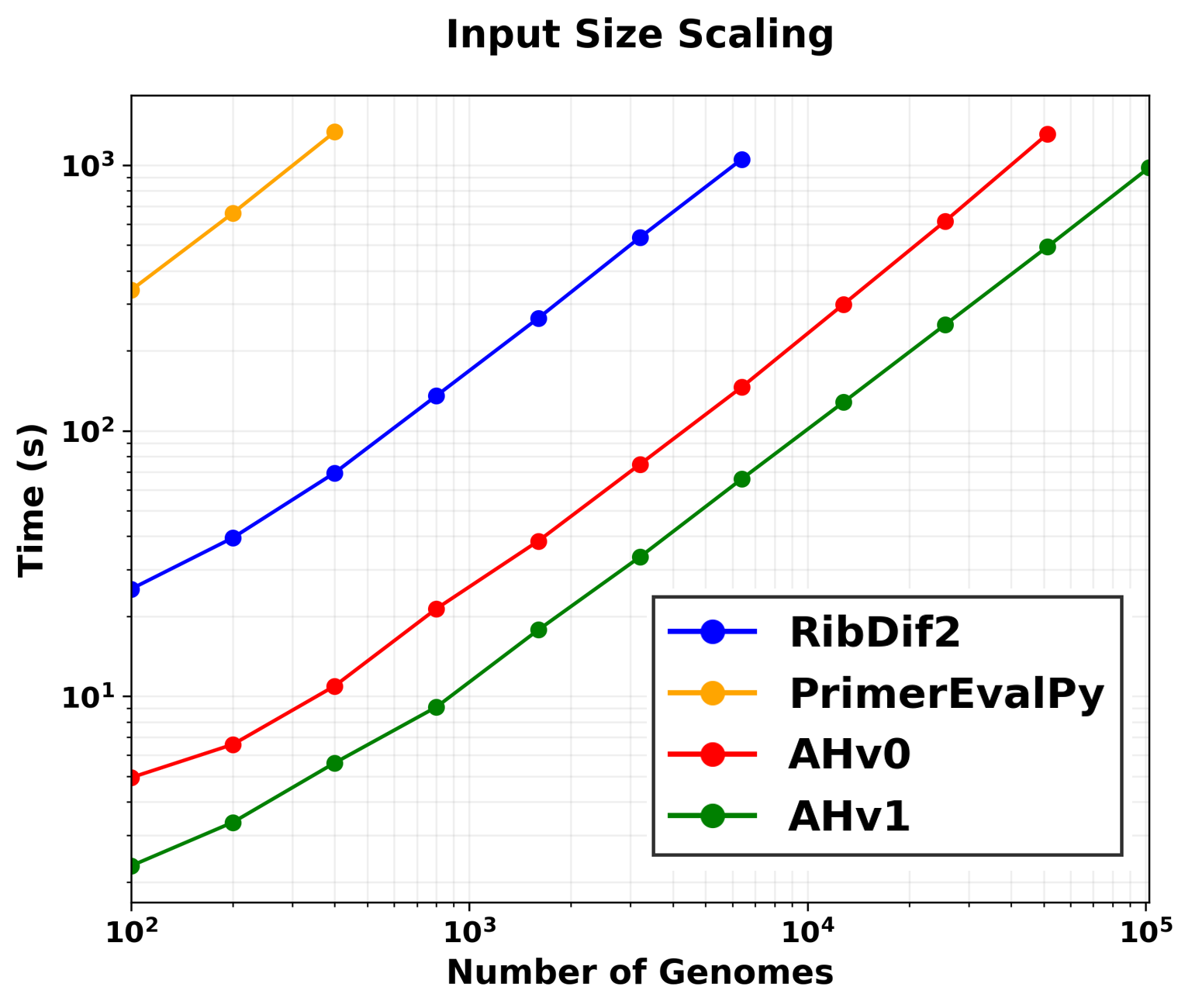}
    \caption{Runtime on subsets of genomes from the AllTheBacteria database \cite{Hunt2024AllTheBacteria} of sizes 100 to 102,400 is shown.}
    \label{fig:input_scaling}
\end{figure}

Figure \ref{fig:input_scaling} shows timing results on datasets ranging in size from 100 to 102,400 genomes. As expected, all methods have near-linear scaling with respect to the number of genomes. AmpliconHunter is two orders of magnitude faster than PrimerEvalPy, one order of magnitude faster than RibDif2, and more than twice faster than the Regex prototype.   AmpliconHunter achieves a throughput of $\sim$100 genomes per second, with the webserver completing the analysis of the full set of 2.4M genomes in 6h 59m 27s.

\comment{
The largest subset of genomes completed by all methods within the allotted time was 400 genomes, on which we took XYZ times.
OR:
The runtime for the largest subset of genomes that each method was able to complete within one hour implies a throughput of 6.07 genomes per second for RibDif2, 0.30 for PrimerEvalPy, 39.2 for AHv0, and 104.7 genomes per second for AHv1.  
}

\subsubsection{Number of mismatches.}
\begin{figure}[tb]
    \centering
    \includegraphics[width=0.5\linewidth]{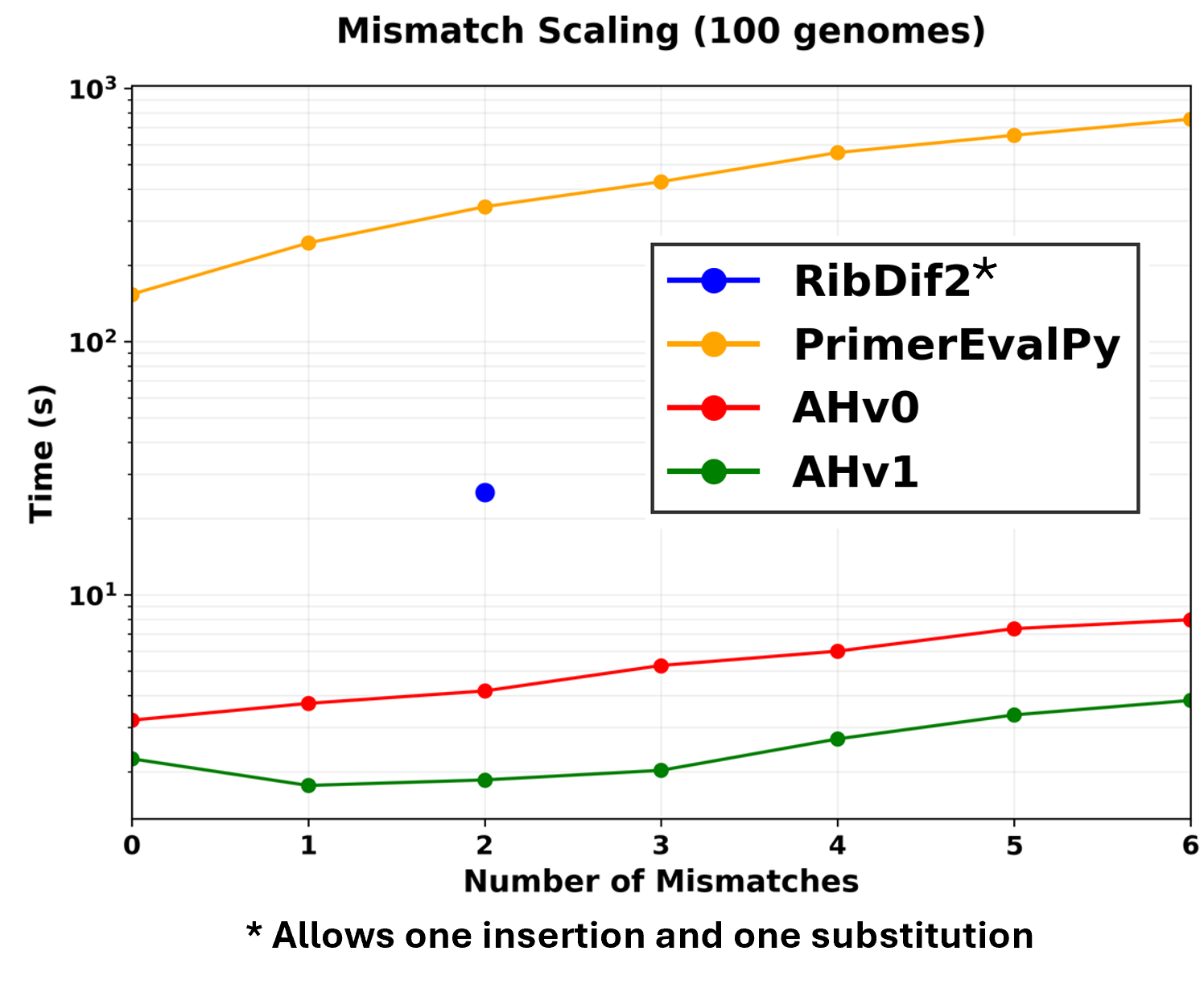}
    \caption{Impact on runtime as number of permitted mismatches is increased. Ribdif2 does not allow alteration of this parameter and thus is shown as a single measurement at $m=2$, as it allows one substitution and one insertion.}
    \label{fig:mismatch_scaling}
\end{figure}

Figure \ref{fig:mismatch_scaling} plots the running time of the compared methods on a fixed dataset of 100 genomes as the maximum number of mismatches is varied from 0 to 6.
The runtime of AmpliconHunter rises moderately with the number of mismatches, from 2.25s to 3.82s for the Hyperscan version, and from 3.19s to 7.97s for the Regex version. In comparison, PrimerEvalPy runtime rises from 16.5s to 89.6s over the same range. While RibDif2 takes 25.3s on the same subset of 100 genomes from the ATB project, it does not expose the number of mismatches as a parameter to be changed. Its primer matching does allow for one indel and one substitution, making it comparable to two mismatches in runtime. At two mismatches, PrimerEvalPy took 340.96s on this dataset, while AHv0 and v1 took 4.17s and 1.86s respectively.

\subsubsection{Primer degeneracy.}
\begin{figure}[tb]
    \centering
    \includegraphics[width=0.5\linewidth]{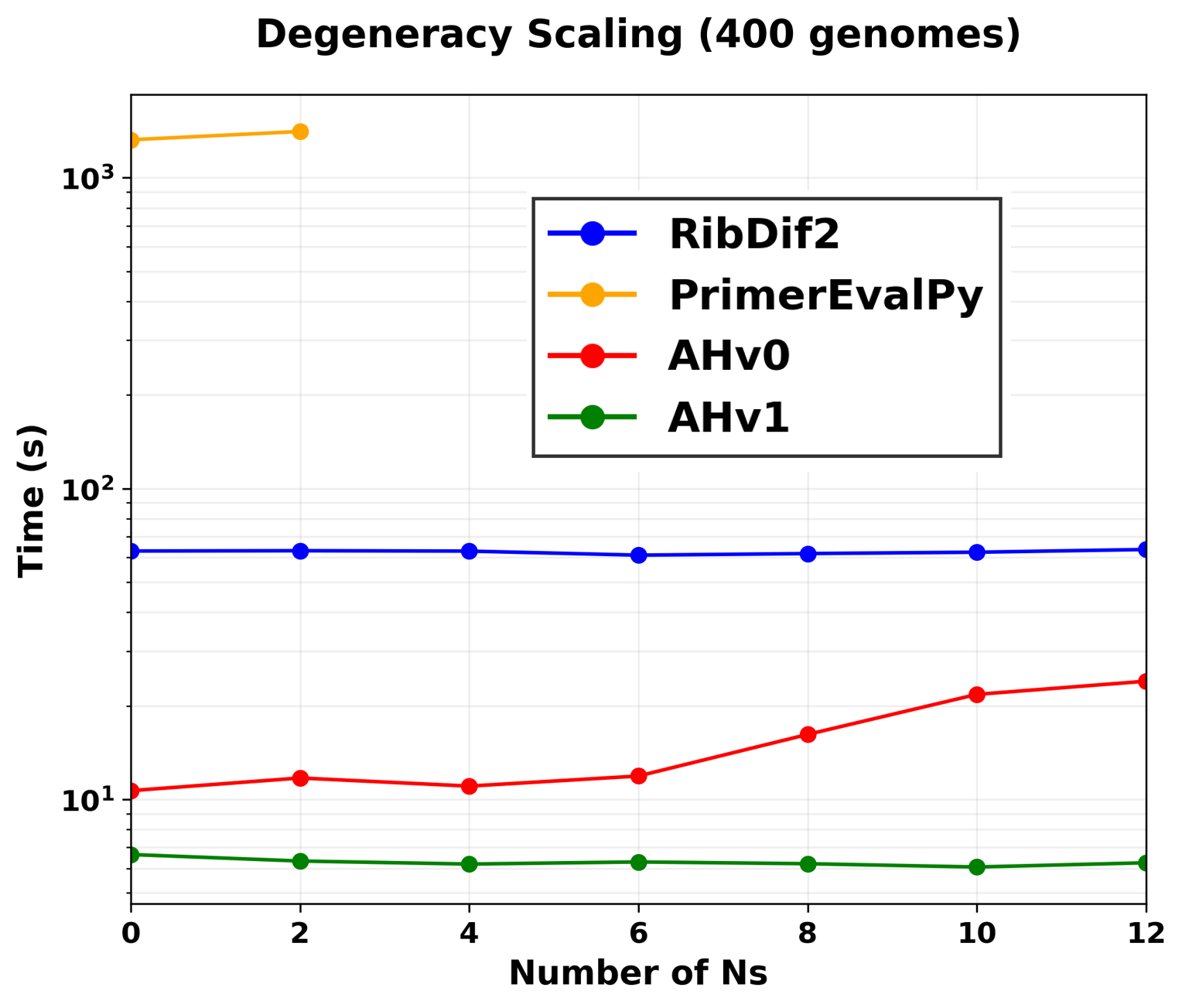}
    \caption{Adding two Ns at a time (one to the 3' end of each primer) increases degeneracy in a controlled manner, and demonstrates its varying impact on runtime between methods.}
    \label{fig:degeneracy_scaling}
\end{figure}

To test the effect of increased degeneracy on the running time we progressively replaced bases from the $3'$ end of the V1V9 primers with fully degenerate bases (`N'). 
Figure \ref{fig:degeneracy_scaling} plots the runtime of the compared methods on 400 random genomes as the number of degenerate bases is increased from 0 to 12 (6 per primer).  Processing times were relatively unchanged for all methods, except for AHv0 which doubles the runtime from 6 to 12 added Ns, from 11.9s to 24.0s.

\subsection{Case Studies}
AmpliconHunter's web interface enables exploration of amplicon patterns across taxonomic groups through several visualizations:

\subsubsection{Amplicon Pattern Visualization}
\begin{figure}[tb]
    \centering
    \includegraphics[width=1\linewidth]{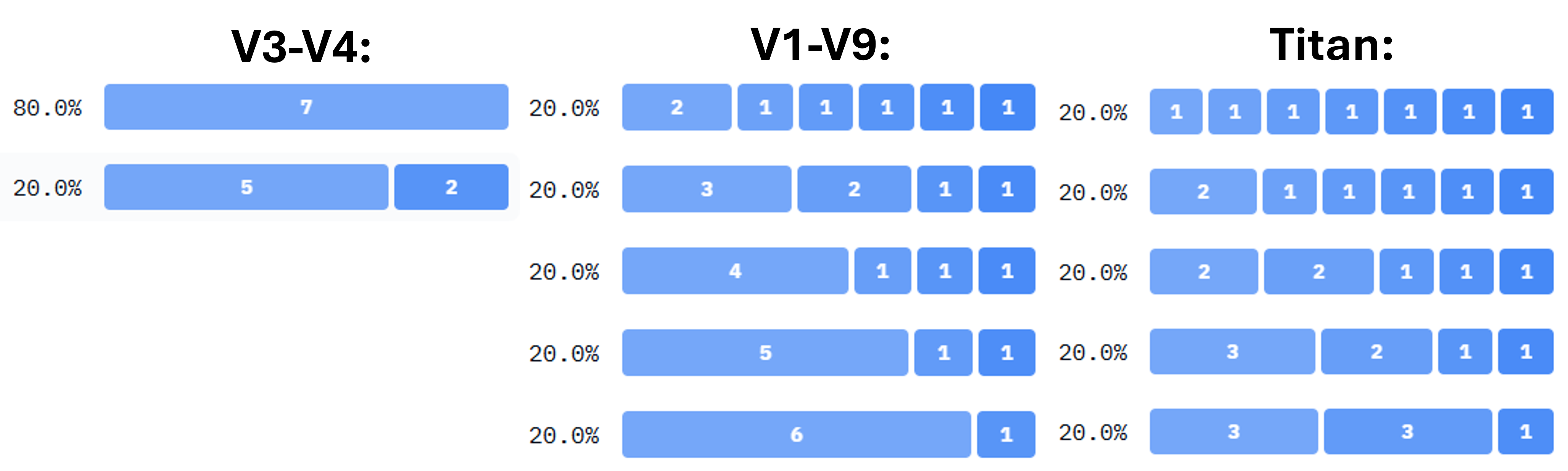}
    \caption{Example amplitype patterns as generated by the web interface - the five complete Escherichia genomes in Refseq generate different amplitypes for three subregions of the rRNA operon: V3-V4, V1-V9, and the Titan\texttrademark \space region.}
    \label{fig:amplitypes}
\end{figure}

The tool identifies distinct amplification patterns within genera and displays them as frequency distributions. For a given genus, each pattern (called an 'amplitype') represents a unique combination of copy numbers and sequence variants. Genomes from the genus Escherichia are known to often have 7 copies of the rRNA operon, which contains the 16S gene \cite{Stoddard2015}. In \ref{fig:amplitypes}, the amplitypes of the five Escherichia genomes in RefSeq-Complete are shown for three subregions of the rRNA operon. In this example, 4 of 5 genomes contain seven identical copies of the V3-V4 region, with one containing two distinct copies in a 5:2 ratio. The amplitypes of V1-V9 for the same genomes exhibit much less sharing, with each genome forming its own unique amplitype, and many more distinct sequences per genome. The Titan region is yet more distinguishing, with every genome containing at least 3 distinct amplicon sequences.

\subsubsection{Species Similarity Analysis}
For each genus, AmpliconHunter can generate heatmaps between all member species showing the pairwise overlap in sets of amplicon sequence variants. This analysis serves multiple purposes. It may help users assess the taxonomic resolution power of different primer pairs within their taxa of interest, in the case of rRNA sequencing. It also allows users to directly examine patterns of sharing for genes of interest between closely related taxa, in the case of antibiotic resistance genes for example.

\begin{figure}[tb]
    \centering
    \includegraphics[width=1.0\linewidth]{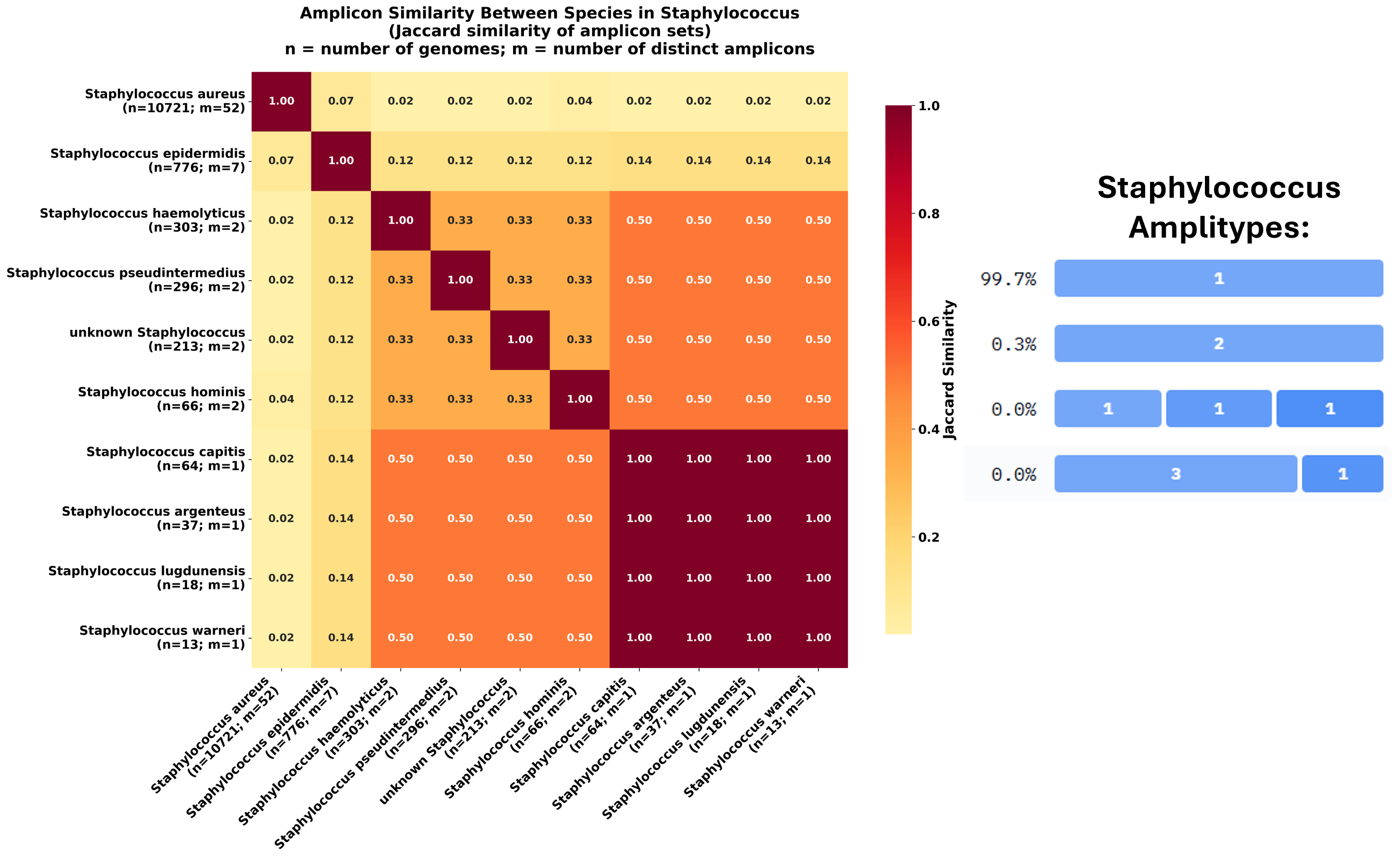}
    \caption{Species similarity heatmap and amplitypes for the 12,557 Staphylococcus genomes the AmpliconHunter web interface identified from GTDB as containing mecA, a known antibiotic resistance gene (forward: GTAGAAATGACTGAACGTCCGATAA, reverse: CCAATTCCACATTGTTTCGGTCTAA) \cite{Hosseini2016,Gorecki2019}. Jaccard similarity between the ten Staphylococcus species with the greatest number of amplified genomes is depicted. For each species, the number of genomes amplified $n$, and the number of distinct amplicons identified $m$ is also shown.}
    \label{fig:mecA_staph}
\end{figure}

\subsubsection{Annealing Temperature Optimization}
Through the web interface, users can explore the relationship between minimum annealing temperature and amplification specificity. Temperature thresholds can be adjusted while monitoring changes in:

\begin{itemize}
    \item Total number of amplicons
    \item Proportion of genomes amplified
    \item Off-target amplification rates
    \item Distribution of patterns of amplicon multiplicity (referred to as `amplitypes')
\end{itemize}

\begin{figure}[tb]
    \centering
    \includegraphics[width=1\linewidth]{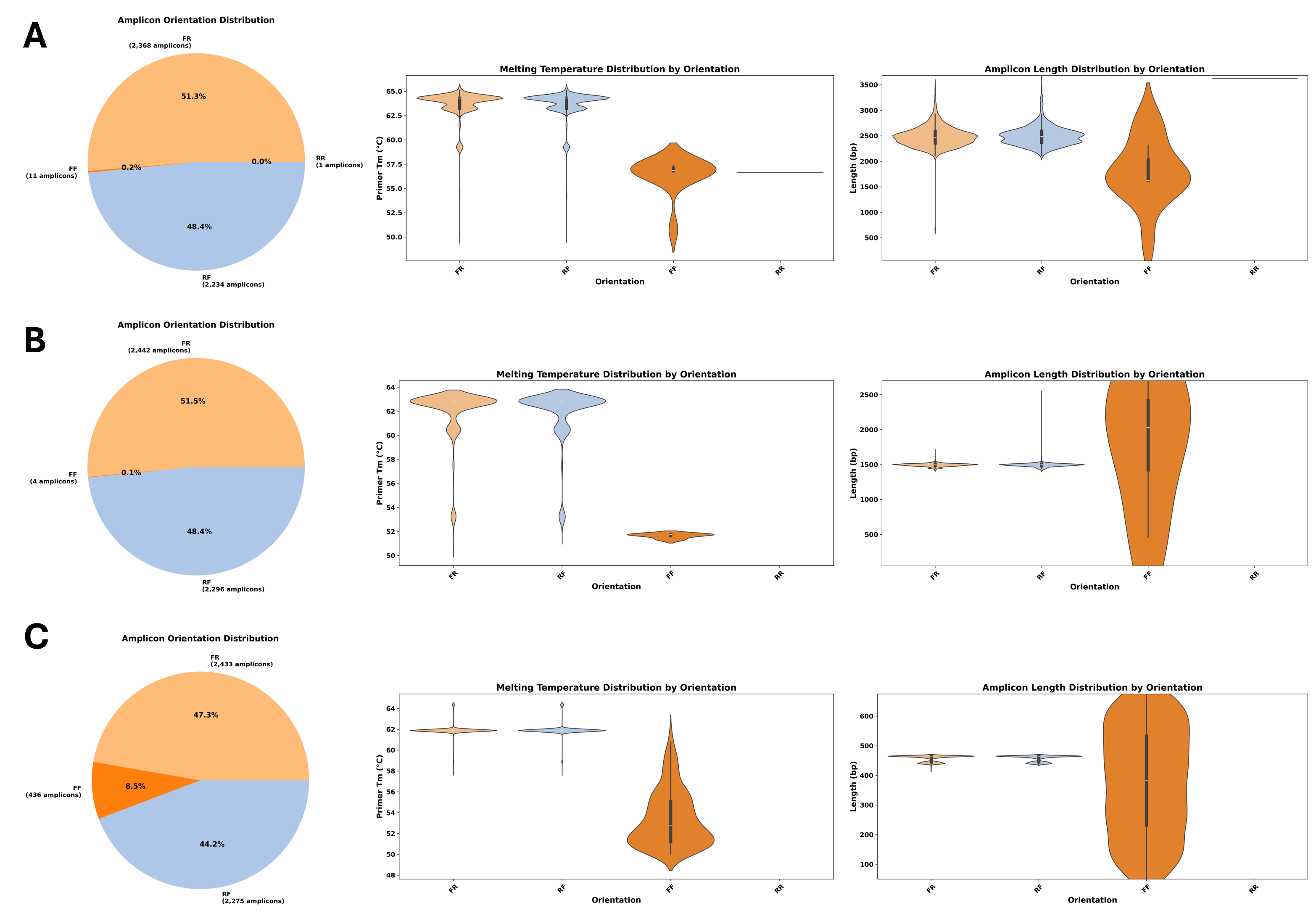}
    \caption{Automatically generated plots of amplicon orientation, melting temperature, and amplicon length distributions for amplicons extracted from the Refseq-complete database with up to six mismatches and melting temperature of 50 using (A) Titan, (B) V1V9, and (C) V3V4 primers.}
    \label{fig:refseq_complete_6_mismatches_autoplots}
\end{figure}

\section{Conclusions and Future Work}
\label{sec.conclusions}

In this paper we introduce AmpliconHunter, a highly scalable \emph{in silico} PCR package distributed as open-source command-line tool and publicly available through a user-friendly web interface at \href{https://ah1.engr.uconn.edu/}{https://ah1.engr.uconn.edu/}.

In all runtime scaling experiments, AmpliconHunter maintained consistent speed advantages over existing tools while providing additional functionality through its temperature-based filtering and taxonomic analysis features.
However, the current implementation searches genomes separately for each pair of primers. As such, genomes must be repeatedly loaded from disk when multiple primer pairs are analyzed. As file input-output represents a significant fraction of the overall runtime, future versions will include functionality to simultaneously analyze multiple primer pairs on a given dataset, dramatically reducing total computational requirements and further improving efficiency.

In ongoing work, we also plan to regularly update and expand the reference databases available through the AmpliconHunter webserver as new genomic assemblies become available. We would like to construct an interface that would allow users to easily search subsets of databases, by taxonomy, as well as assembly characteristics in the case of RefSeq. We also plan to add more visualizations and interactivity. Specifically, we would like to leverage the profile HMMs that are already built to automatically align and build trees for amplicon sequences with the same amplitype to analyze sequence similarity within an amplitype. This could be used to identify signatures of horizontal gene transfer in reference databases. We also plan to leverage AmpliconHunter to create databases for machine learning methods to cluster amplicon sequences according to their genome of origin.

\bibliography{references}

\newpage
\appendix
\section{Evidence of Off-Target Amplification in Experimental Data}
\label{appendix:offtarget}

\begin{figure}[tb]
    \centering
    \includegraphics[width=1.0\linewidth]{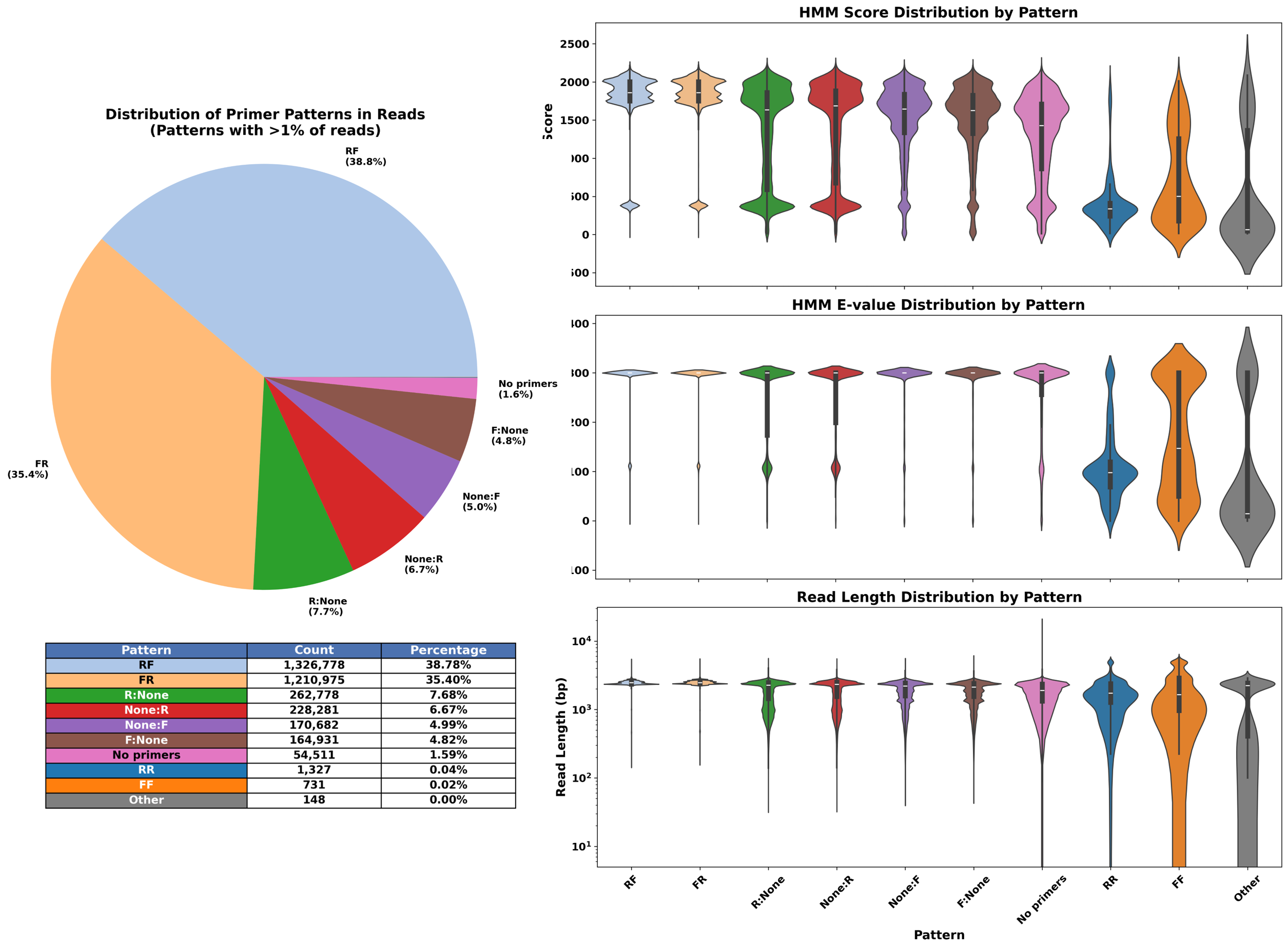}
    \caption{True primer orientation, as measured using real Titan reads generated via PacBio CCS long-read technology. Patterns are determined through regex matching of primer sequences on the first and last 100BPs of each read, allowing up to two substitutions. No melting temperature or clamp filtering was applied. HMM scoring was conducted using an HMM produced from running AmpliconHunter on the RefSeq-complete database with default parameters.}
    \label{fig:taconic_primer_orientation}
\end{figure}

\begin{figure}[tb]
    \centering
    \includegraphics[width=0.85\linewidth]{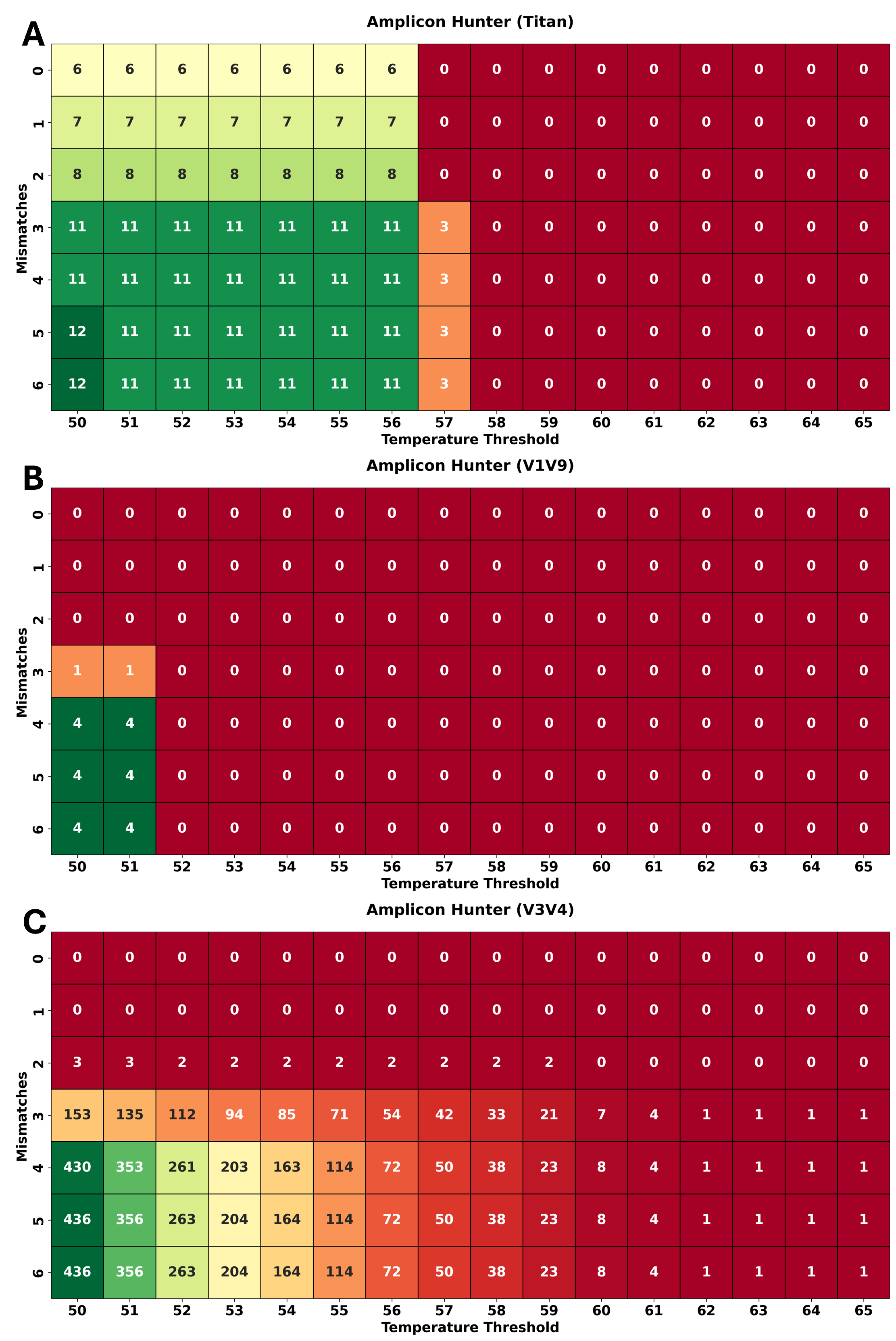}
    \caption{A heatmap depicting the number of predicted off-target amplicons filtered from accuracy experiments on a subset of the RefSeq-Complete database using primer pairs targeting three different regions: Titan, V1V9, and V3V4.}
    \label{fig:offtarget_filtered_pred}
\end{figure}

Off-target amplification is shown to be a small but detectable proportion of output of experimentally obtained PacBio CCS long reads using Titan primers in \autoref{fig:taconic_primer_orientation}. However, it is possible that less diligently designed primer pairs may lead to greater incidence of this phenomenon. In our previous accuracy experiments, we filtered out the off-target amplicons generated for the three primer pairs run on the RefSeq-Complete database with various parameter settings, prior to calculating the F1 score. Despite their moderate level of degeneracy, V3V4 primers are predicted to exhibit the highest off-target amplification, whereas the highly degenerate Titan primers show minimal off-target amplification (\autoref{fig:offtarget_filtered_pred}).

\end{document}